\documentclass[aps,reprint,groupedaddress, amsmath, amssymb]{revtex4-1}
\usepackage{graphicx}
\usepackage{bm}
\usepackage{amsmath}
\usepackage{amssymb}
\usepackage{float}
\usepackage{multirow}

\begin{document}
\title{Electrostatic Point Charge Fitting as an Inverse Problem: \\ Revealing the Underlying Ill-Conditioning}

\author{Maxim V. Ivanov}
\author{Marat R. Talipov}
\author{Qadir K. Timerghazin}
\email{qadir.timerghazin@marquette.edu}
\affiliation{Department of Chemistry, Marquette University, Milwaukee, Wisconsin}

\date{\today}

\begin{abstract}
Atom-centered point charge model of the molecular electrostatics---a major workhorse of the atomistic biomolecular simulations---is usually parameterized by least-squares (LS) fitting of the point charge values to a reference electrostatic potential, a procedure that suffers from numerical instabilities due to the ill-conditioned nature of the LS problem. 
To reveal the origins of this ill-conditioning, we start with a general treatment of the point charge fitting problem as an inverse problem, and construct an analytical model with the point charges spherically arranged according to Lebedev quadrature which is naturally suited for the inverse electrostatic problem. 
This analytical model is contrasted to the atom-centered point-charge model that can be viewed as an irregular quadrature poorly suited for the problem. 
This analysis shows that the numerical problems of the point charge fitting are due to the decay of the curvatures corresponding to the eigenvectors of LS sum Hessian matrix. 
In part, this ill-conditioning is intrinsic to the problem and related to decreasing electrostatic contribution of the higher multipole moments, that are, in the case of Lebedev grid model, directly associated with the Hessian eigenvectors.
 For the atom-centered model, this association breaks down beyond the first few eigenvectors related to the high-curvature monopole and dipole terms; this leads to even wider spread-out of the Hessian curvature values. 
Using these insights, it is possible to alleviate the ill-conditioning of the LS point-charge fitting without introducing external restraints and/or constraints. 
Also, as the analytical Lebedev grid PC model proposed here can reproduce multipole moments up to a given rank, it may provide a promising alternative to including explicit multipole terms in a force field. 

\end{abstract}
\maketitle
\section{Introduction}
The atom-centered point charge (PC) model of molecular electrostatics has been a mainstay of biomolecular simulations for decades.\cite{Cox1981, Singh1984, Bayly1993, Hinsen1997, Simmonett2005,Dupradeau2010,Huang2013, Gotz2014, Dickson2014, Arnautova2015, Sprenger2015, Mukhopadhyay2015, Maier2015, Gabrieli2015}
While chemically intuitive and straightforward in technical implementation, this model does not provide a sufficiently detailed description of the anisotropic features of the molecular electrostatic potential (MEP), such as lone pairs, $\pi$-systems, and $\sigma$-holes, etc. which are mostly governed by higher-order multipole terms.\cite{Cardamone2014,Kramer2014}
These anisotropic effects, however, can be described within the PC approximation by moving beyond the atom-centered paradigm, i.e. by adding non-atom centered PCs/extended points.\cite{Karamertzanis2004, Cerutti2013, Gao2014, Devereux2014} 
Although increasing the number of PCs per atom improves the quality of the electrostatic model, it also can exacerbate well-known ill-conditioning and redundancy problems\cite{Francl1995, Sigfridsson1997, Jakobsen2014} of the PC fitting procedures, leading to numerically unstable solutions.\cite{Tschampel2007, Dixon1997, Bayly1993}

These numerical instabilities are usually related to a large variation of the PC values for atoms in the interior of the molecule, so-called buried atom effect.\cite{Stouch1992,Stouch1993,Bayly1993, Hinsen1997} 
The buried atom (usually methyl and methylene carbons) charges can dramatically change due to trivial changes in the PC fitting problem (the probe grid sampling, spatial orientation of the molecule, etc.), and/or have  inconsistent values across very similar molecules or even conformers of the same molecule.\cite{Breneman1990,Woods1989}
As the inclusion of non-atom centered PCs into the model produces even more buried centers, it should also increase the numerical instabilities of the PC fitting problem.

In fact, these numerical problems are rooted in the mathematical nature of the PC derivation---the least squares (LS) fitting to the reference MEP:\cite{Cox1981, Singh1984}
\begin{equation} \label{eq:chi2}
\chi^2 \left( \mathbf{q} \right) = | \mathbf{\Phi} - \mathbf{A}\mathbf{q} |^2 = |\mathbf{\Phi}|^2 + \mathbf{g}^\intercal \cdot \mathbf{q} + \mathbf{q}^\intercal \mathbf{H} \mathbf{q},
\end{equation}
\begin{equation}
\mathbf{g} = -2\mathbf{A}^\intercal \mathbf{\Phi},
\end{equation}
\begin{equation} \label{eq:hessian}
\mathbf{H} = \mathbf{A}^\intercal \mathbf{A},
\end{equation}
where the LS sum $\chi^2$ is the subject of minimization and the solution satisfies normal equations:\cite{lawson74}
\begin{equation} \label{eq:normal}
    \mathbf{A}^\intercal \mathbf{A} \mathbf{q} = \mathbf{A}^\intercal \mathbf{\Phi}.
\end{equation}
Here,
the elements of the LS matrix $\mathbf{A}$ correspond to the inverse distance $1/r_{ij}$ 
between the PC $i$ and the grid point $j$;
$\mathbf{\Phi}$ is $T$-dimensional vector of the reference values of MEP;
$\mathbf{q}$ is $N$-dimensional vector of the PC values;
$\mathbf{g}$ is the gradient of the function $\chi^2$ at the origin ($\mathbf{q}=0$); 
$\mathbf{H}$ is the Hessian matrix of LS sum $\chi^2$.

While the ill-conditioning is common to many LS fitting problems,\cite{Hansen2010, Machta2013, Transtrum2015}
numerical difficulties associated with PC fitting are further compounded by commonly used total charge constraint using Lagrange multiplier.\cite{Besler1990,Francl1995, Cieplak1995,Sigfridsson1997}

One of the most widely used techniques to alleviate the numerical instabilities of PC fitting is to add artificial restraints to the PC values of the buried atoms.\cite{Bayly1993, Dupradeau2010, Zeng2013,Burger2013} Although this method can be extended to models with off-center PCs/extended points, one may wonder if it would be possible to overcome these difficulties in a more elegant way, based on better physical understanding of the problem. 

For instance, an important insight can be gleaned from the eigendecomposition of the LS sum Hessian matrix (eq. \ref{eq:hessian}):
\begin{equation} \label{eq:kappa}
\mathbf{H} \mathbf{u}_i = \kappa_i \mathbf{u}_i.
\end{equation}
Indeed, the ill-conditioned nature of the LS matrix $\mathbf{A}$ can be related to the significant differences in the eigenvalues $\kappa_i$, i.e. the LS sum curvatures along the directions defined by the eigenvectors $\mathbf{u}_i$.\cite{Ivanov2015,Hansen2013} Because of the 2--3 order of magnitude variation of the $\kappa_i$ values, different sets of PCs can produce essentially the same MEP, as these solutions have the same positions along the high-curvature directions, although the positions along the low-curvature directions could be quite different.\cite{Ivanov2015} Importantly, the eigenvectors with the largest curvatures usually correspond to the total charge and dipole moment components of the molecule, while the lower-curvature eigenvectors do not seem to be associated with particular multipole moments.\cite{Ivanov2015, Laio2002}

However, the exact physical origin of the correspondence between the large curvature eigenvectors and the first terms of the multipole expansion is unclear, along with the nature of the low-curvature eigenvectors.
Particularly, it is not clear if the presence of the low-curvature modes of the $\mathbf{H}$ matrix and thus the ill-conditioning of the LS problem is solely because of the nature of the PC fitting problem, or due to some numerical factors, e.g. an incomplete sampling of the reference MEP grid.

To address these questions, we here revisit the PC fitting problem from the first principles. While the atom-centered PC model traces back to the intuitive chemical concept of the atomic charge, we consider a general PC model as a case of the inverse problem, where one seeks to recover the source charge distribution from its effect, i.e. electrostatic potential distribution. Based on the properties of the Coulomb law, we construct a best-case electrostatic model for which the inverse problem can be solved exactly, both in the continuous case, as well as in the case of a discrete (non-atom centered) PC approximation. 

Using this model, we investigate the nature of the eigenvectors $\mathbf{u}_i$ and their eigenvalues $\kappa_i$, and dissect the factors responsible for the ill-conditioning of the LS fitting problem, and discuss how these insights can be used to improve and simplify the existing PC derivation procedures.

\section{Point Charge Fitting as an Inverse Problem}
A problem where given an effect (in this case the MEP $\Phi$) defined in the region $V_\Phi$, its cause (a charge distribution $\rho$) defined in the region $V_\rho$ needs to be determined belongs to a general class of inverse problems and can be described by the Fredholm integral equation of the first kind:\cite{Groetsch2007}
\begin{equation}\label{eq:integral}
    \int_{V_\rho} k(\mathbf{r}, \mathbf{r'})  \rho \left( \mathbf{r}'  \right) d \mathbf{r'} = \Phi \left( \mathbf{r} \right),
\end{equation}
where kernel $k(\mathbf{r}, \mathbf{r'})$ 
specifies the evolution of the cause $\rho(\mathbf{r}')$ into the effect  $\Phi(\mathbf{r})$, that in this case corresponds to the Coulomb law:
\begin{equation} \label{eq:kernel}
    k(\mathbf{r}, \mathbf{r}') = \frac{1}{|\mathbf{r}-\mathbf{r'}|}.
\end{equation}

The integral equation can also be represented as an operator equation:
\begin{equation} \label{eq:operator}
    K\rho=\Phi,
\end{equation}
where $K:U \rightarrow V$ is a linear operator defined on space $U=\mathrm{range}(K^\ast) \in L^2$ of square integrable functions,
and takes values in space $V=\mathrm{range}(K) \in L^2$; 
$K^\ast: V \rightarrow U$ is adjoint of $K$. This equation can be solved exactly if and only if $\Phi \in V$. 
However, in general it is not the case, so a function $\rho$ that minimizes the residual norm $| \Phi-K\rho |$ 
is considered as the LS solution and thus satisfies the normal equation:\cite{Groetsch2007, Roman2008}
\begin{equation}
    K^\ast K \rho = K^\ast \Phi.
\end{equation}

This LS solution can be obtained as the linear combination of the basis vectors $u_i \in U$: \cite{Groetsch2007}
\begin{equation} \label{eq:inverse}
  \rho = K^\dagger \Phi=\sum_{i=1}^\infty \frac{\langle\Phi,v_i\rangle}{\mu_i}u_i,
\end{equation}
where $K^\dagger$ is the Moore-Penrose inverse, $\mu_i$ is a singular value, $v_i$ and $u_i$ are left and right singular vectors, respectively and the inner product $\langle\Phi,v_i\rangle$ is defined as
\begin{equation}
	\langle\Phi,v_i\rangle = \int_{V_\Phi} \Phi(\mathbf{r}) v_i(\mathbf{r}) d \mathbf{r}
\end{equation}

The orthogonal bases $\{u_i\}_{i=1}^\infty$ and $\{v_i\}_{i=1}^\infty$ also form the eigenbases of $K^\ast K$ and $K K^\ast$ with eigenvalues $\mu_i^2$:
\begin{equation} \label{eq:K*K}
K^\ast K u_i = \mu_i^2 u_i,
\end{equation}
\begin{equation} \label{eq:KK*}
K K^\ast v_i = \mu_i^2 v_i.
\end{equation}

To obtain a numerical solution to the integral equation (eq. \ref{eq:integral}), the regions over which the MEP and charge distribution are defined are sampled using a numerical quadrature. Given $N$ quadrature nodes over the charge distribution and $T$ nodes over the MEP region the integral equation is transformed into a system of $T$ linear equations:
\begin{equation} \label{eq:linear}
    \mathbf{K} \mathbf{q} = \mathbf{\Phi}, 
\end{equation}
where the $T\times N$ matrix $\mathbf{K}$ is identical to the LS matrix $\mathbf{A}$ from eq. \ref{eq:chi2} and contains the kernel elements $k_{ij}$, as this matrix originates from the kernel $k(\mathbf{r}, \mathbf{r'})$ in the integral equation (eq. \ref{eq:integral}). It will be further referred to as $\mathbf{K}$ in order to highlight its mathematical origin.

Then, the PC value at the node $i$ is the product of the charge density $\rho_i$ and the quadrature weight $w_i$:
\begin{equation}
	q_i = \rho_i w_i
\end{equation}

Since the number of the reference values $T$ is usually larger than the number of the unknown PC values $N$, the system of linear equations is overdetermined. Then, a solution that minimizes the LS sum $\chi^2 \left( \mathbf{q} \right)$ (eq. \ref{eq:chi2}) and satisfies normal equations (eq. \ref{eq:normal}) is considered as the numerical solution to the integral equation (eq. \ref{eq:integral}). This solution can be obtained using singular value decomposition (SVD) of matrix $\mathbf{K}$: \cite{lawson74, Hansen2013, Roman2008}
\begin{equation} \label{eq:svd}
  \mathbf{q} = K^\dagger \mathbf{\Phi}=\sum_{i=1}^{r} \frac{ \mathbf{\Phi} \cdot \mathbf{v}_i } {\mu_i} \mathbf{u}_i,
\end{equation}
where  $K^\dagger$ is the Moore-Penrose pseudoinverse;
$\mu_i$ are singular values of matrix $\mathbf{K}$;
vectors $\mathbf{v}_i$ and $\mathbf{u}_i$ are left and right singular vectors.
If the rank $r$ of matrix $\mathbf{K}$ is less than the dimension of $\mathbf{q}$ ($r<N$), then the matrix $\mathbf{K}$ is rank deficient.

Similarly to the continuous case (eqs. \ref{eq:K*K}-\ref{eq:KK*}), the orthogonal bases $\{v_i\}_{i=0}^r$ and $\{u_i\}_{i=0}^r$ form eigenbases for $\mathbf{K} \mathbf{K}^\intercal$ and $\mathbf{K}^\intercal \mathbf{K}$:
\begin{equation}
    \mathbf{K} \mathbf{K}^\intercal \mathbf{v}_i = \mu_i^2 \mathbf{v}_i,
\end{equation}
\begin{equation} \label{eq:curvature}
    \mathbf{K}^\intercal \mathbf{K} \mathbf{u}_i = \mu_i^2 \mathbf{u}_i,
\end{equation}
where $\mathbf{K}^\intercal \mathbf{K}$ is also a Hessian matrix (eq. \ref{eq:kappa}) 
and $\mu_i^2$ is identical to its eigenvalue $\kappa_i$, which is the $\chi^2$ curvature along the direction $\mathbf{u}_i$:\cite{Brown2003}
\begin{equation}
	 \mu_i^2 =\kappa_i
\end{equation}

In many LS problems, PC fitting included, the singular values vary in a wide range, revealing the underlying ill-conditioning.\cite{Francl1995, Sigfridsson1997, Hansen2010, Machta2013} 
As a singular value $\mu_i$ is a denominator in the LS solution (eq. \ref{eq:svd}), the smaller the singular value, the larger the effect of the corresponding singular vector $\mathbf{u}_i$ on the LS solution. Thus, even small variations along $\mathbf{u}_i$ with small singular value lead to a significant variations of the LS solution, although these variations do not lead to significant change in the quality of the fit $\chi^2$.\cite{Ivanov2015}
To understand the origins of the ill-conditioning in PC fitting, we next consider a system for which the inverse electrostatic problem can be analytically solved.

\begin{figure}[H]
\includegraphics{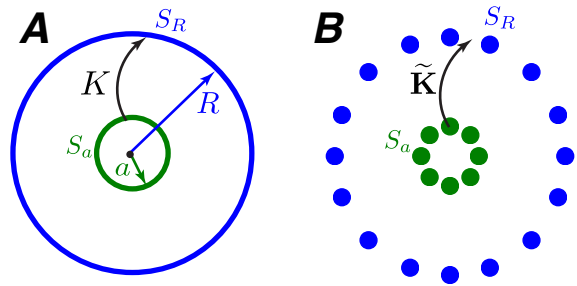}
\caption{\label{fig:twospheres} 
Schematic representations of the probe $S_R$ and charged $S_a$ spheres in the continuous (A) and discrete (B) forms.
Operators $K$ (eq. \ref{eq:K}) and matrix $\widetilde{\mathbf{K}}$  
are represented schematically.}
\end{figure}

\section{The Two-Sphere Model}

The Coulomb kernel (eq. \ref{eq:kernel}) can be conveniently expanded in terms of spherical harmonics so the source $\mathbf{r}'$ and the observation $\mathbf{r}$ coordinates are separated but share the same origin:\cite{Stone,Jackson1999}
\begin{equation} \label{eq:expansion}
\begin{gathered}
k(\mathbf{r}, \mathbf{r}') = \frac{1}{|\mathbf{r}-\mathbf{r'}|} \\ 
= \sum_{l=0}^\infty \sum_{m=-l}^l \frac{4\pi}{2l+1} \frac{r_<^l}{r_>^{l+1}} Y_{lm} (\widehat{ \mathbf{r}'}) Y_{lm} (\widehat{ \mathbf{r} }),
\end{gathered}
\end{equation}
where $\widehat{ \mathbf{r}} = \mathbf{r}/r$ denotes the unit vector 
defined by the polar $\varphi$ and azimuthal $\theta$ angles;
$r_<$ is the smaller and $r_>$ is the larger of $r$ and $r'$;
$Y_{lm}$ are orthogonal real-value spherical harmonics:\cite{f1}
\begin{equation} \label{eq:ortho}
    \int_S Y_{lm}(\widehat{\mathbf{r}}) Y_{l'm'}(\widehat{\mathbf{r}}) d \Omega = \delta_{ll'} \delta_{mm'},
\end{equation}
where $d \Omega$ is the differential of the solid angle.

Then, in the region beyond the divergence sphere where the charge density vanishes, 
the MEP can be expanded in a multipole series:\cite{Jackson1999, Stone}
\begin{equation} \label{eq:mexp}
    \Phi(\mathbf{r}) = \sum_{l=0}^\infty \sum_{m=-l}^l \sqrt{ \frac{4\pi}{2l+1}  } r^{-l-1}  Q_{lm}^{mol} Y_{lm} (\widehat{ \mathbf{r}  }),
\end{equation}
where a molecular multipole moment $Q_{lm}^{mol}$ is given by
\begin{equation} \label{eq:Qlm}
    Q_{lm}^{mol} = \sqrt{ \frac{4\pi}{2l+1} } \int r^l \rho(\mathbf{r}) Y_{lm} (\widehat{\mathbf{r}}) d^3r.
\end{equation}

The form of the kernel expansion (eq.\ref{eq:expansion}) suggests that if the radii $r=R$ and $r'=a$ are fixed, the kernel $k(\mathbf{R}, \mathbf{a})$ can uniquely map a charge density over a spherical surface $S_a$ to the corresponding potential $\Phi(\mathbf{R})$ on a sphere $S_R$ and vice versa. 
Thus, for a probe sphere $S_R$ with the radius $R$ greater than the radius of divergence sphere, the MEP can be reproduced exactly by a sphere $S_a$ with surface charge density $\sigma(\mathbf{a})$ such that the multipole moments of the sphere $Q_{lm}^{S_a}$ are equivalent to the multipole moments of the molecule $Q_{lm}^{mol}$:
\begin{equation} \label{eq:equiv}
    Q_{lm}^{S_a} \equiv Q_{lm}^{mol},
\end{equation}
where multipole moments of the sphere are:
\begin{equation}\label{eq:Qlm-sphere}
    Q_{lm}^{S_a} = \sqrt{ \frac{4\pi}{2l+1} } a^l \int_{S_a} \sigma(\mathbf{a}) Y_{lm}(\widehat{\mathbf{a}}) d\Omega.
\end{equation}

In this case, the original integral eq. \ref{eq:integral} is transformed into a surface integral equation:
\begin{equation} \label{eq:surface}
  \int_{S_a} k(\mathbf{R}, \mathbf{a}) \sigma(\mathbf{a}) d\Omega = \Phi(\mathbf{R}),
\end{equation}
or, equivalently, in an operator form
\begin{equation}
    K \sigma = \Phi,
\end{equation}
where $K: L^2(S_a) \rightarrow L^2(S_R)$ is a compact infinite-rank operator (Fig. \ref{fig:twospheres}A):
\begin{equation} \label{eq:K}
    K \sigma = \sum_{l=0}^\infty \sum_{m=-l}^l \mu_l \langle \sigma, Y_{lm}^{S_a} \rangle Y_{lm}^{S_R}, 
\end{equation}
where subscripts $S_a$ and $S_R$ denote the spheres, on which the corresponding spherical harmonics are defined;
the projection $\langle \sigma, Y_{lm}^{S_a}  \rangle$ is the inner product on the $L^2(S_a)$ space:
\begin{equation} \label{eq:sigmaY}
    \langle \sigma, Y_{lm}^{S_a}  \rangle = \int_{S_a} \sigma(\mathbf{a}) Y_{lm}^{S_a}(\widehat{\mathbf{a}}) d \Omega
\end{equation}
and for each degree $l$  there is a singular value $\mu_l$ in the form of the distance-dependent factor from the MEP expansion (eq. \ref{eq:expansion}):
\begin{equation} \label{eq:mu}
\mu_l = \frac{4\pi}{2l+1} \frac{a^l}{R^{l+1}}.
\end{equation}

Accordingly, the spherical harmonics $Y_{lm}^{S_R}$ and $Y_{lm}^{S_a}$ are left and right singular vectors and thus the eigenfunctions of the operators $K^\ast K$ and $K K^\ast$, while the squares of the singular values $\mu_l$ are their eigenvalues (eqs. \ref{eq:K*K}-\ref{eq:KK*}).
Since the singular values $\mu_l$ and spherical harmonics $Y_{lm}^{S_a}$ and $Y_{lm}^{S_R}$ form a singular system of the operator $K$, the solution to integral equation (eq. \ref{eq:surface}) can be expressed as: 
\begin{equation} \label{eq:K+}
    \sigma = K^\dagger \Phi = \sum_{l=0}^\infty \sum_{m=-l}^l \frac{\langle \Phi, Y_{lm}^{S_R}  \rangle}{\mu_l} Y_{lm}^{S_a}.
\end{equation}

According to the multipole expansion (eq. \ref{eq:mexp}), the inner  product $\langle \Phi, Y_{lm}^{S_R} \rangle$ depends on the radius $R$ of the probe sphere and the multipole moments of the molecule:
\begin{equation} \label{eq:phiY}
    \langle \Phi, Y_{lm}^{S_R}  \rangle = \int_{S_R} \Phi(\mathbf{R}) Y_{lm}^{S_R}(\widehat{\mathbf{R}}) d \Omega = 
    \sqrt{\frac{4\pi}{2l+1}} \frac{1}{R^{l+1}}Q_{lm}^{mol}.
\end{equation}

The dependence on the radius $R$ cancels out, so the charge density depends only on the radius $a$ of the sphere $S_a$ and the molecular multipole moments:
\begin{equation} \label{eq:sigma}
    \sigma(\mathbf{a}) = \sum_{l=0}^\infty \sum_{m=-l}^{l} \sqrt{ \frac{2l+1}{4\pi} } a^{-l}
    Y_{lm}^{S_a}(\widehat{\mathbf{a}})Q_{lm}^{mol},
\end{equation}
and the charged sphere $S_a$ exactly reproduces the MEP $\Phi(\mathbf{R})$.

\section{Analytical Lebedev Grid Point Charge Model}

We can construct an approximate discrete analog of the two-sphere model (eqs. \ref{eq:surface}-\ref{eq:sigma}, Fig. \ref{fig:twospheres}) using a quadrature that exactly integrates spherical harmonics $Y_{lm}$ over a sphere up to a given $l$ (eqs. \ref{eq:sigmaY} and \ref{eq:phiY}), e.g. the widely used\cite{Leang2012,Steinmann2012,Parrish2013} Lebedev quadrature,\cite{Lebedev1999} that defines $N$ quadrature nodes (Table I in the supplementary material\cite{SI}) with predetermined angular coordinates $\theta_i$, $\varphi_i$, and integration weights $w_i$:
\begin{equation}
    \int_S Y_{lm}(\theta, \varphi) d \Omega = \sum_i^N Y_{lm}(\theta_i, \varphi_i) w_i.
\end{equation}
Then, given the surface charge density $\sigma_i$ the corresponding point charge is:
\begin{equation}
	q_i = \sigma_i w_i.
\end{equation}

Due to the orthogonality of the spherical harmonics $Y_{lm}$ (eq. \ref{eq:ortho}), the $N$-node Lebedev quadrature that exactly integrates spherical harmonics over the sphere $S_a$ up to $l=2n$
\begin{equation} \label{eq:ortho2}
    \sum_i^N Y_{lm}^{S_a}(\theta_i, \varphi_i) 
    Y_{l'm'}^{S_a}(\theta_i, \varphi_i) w_i^{S_a} = 
    \widetilde{\mathbf{Y}}_{lm}^{S_a} \cdot \widetilde{\mathbf{Y}}_{l'm'}^{S_a} = 
    \delta_{ll'} \delta_{mm'},
\end{equation}
defines an orthonormal basis:
\begin{equation} \label{eq:Ybasis}
    \widetilde{\mathbf{Y}}_{S_a} = \{\ \widetilde{\mathbf{Y}}_{lm}^{S_a},\ -l\leq m \leq l\ \}_{l=0}^n,
\end{equation}
where the $\widetilde{\mathbf{Y}}_{lm}^{S_a}$ vectors have $d_n = (n+1)^2$ elements defined as:
\begin{equation}
    \widetilde{Y}_{lmi}^{S_a} = Y_{lm}(\theta_i, \varphi_i) \sqrt{w^{S_a}_i}.
\end{equation}
Similarly, the probe sphere $S_R$ can be represented by a $T$-node Lebedev grid that integrates spherical harmonics up to $l=2t$ and  defines an orthogonal basis $\widetilde{\mathbf{Y}}_{S_R}$ of dimension $d_t = (t+1)^2$.

In this discrete representation, the operator $K$ (eq. \ref{eq:K}) then becomes a $T \times N$ matrix $\widetilde{\mathbf{K}}$:\cite{Ahrens2009} 
\begin{equation} \label{eq:Ktilde}
    \widetilde{\mathbf{K}} \ \widetilde{\pmb{\sigma}}\ = \widetilde{\mathbf{\Phi}},
\end{equation}
where the elements of $\widetilde{\mathbf{K}}$, $\widetilde{\pmb{\sigma}}$, and $\widetilde{\mathbf{\Phi}}$ are:
\begin{gather}
    \widetilde{K}_{ij} = \sqrt{w_i^{S_a} w_j^{S_R}} / r_{ij},\\
    \widetilde{\sigma}_i = \sigma_i \sqrt{w_i^{S_a}},\ \ 
    \widetilde{\Phi}_j = \Phi_j \sqrt{w_j^{S_R}}.
\end{gather}

Since usually the probe grid has more points than the source grid, i.e. $T>N$, the matrix equation (eq. \ref{eq:Ktilde}) is a LS problem (eq. \ref{eq:chi2}) that can be solved using SVD of the matrix $\widetilde{\mathbf{K}}$ (eq. \ref{eq:svd}), giving a discrete analog of eq. \ref{eq:sigma}:
\begin{equation} \label{eq:svd2}
    \widetilde{\pmb{\sigma}} = \sum_{l=0}^n \sum_{m=-l}^l 
    \frac{\widetilde{\mathbf{\Phi}} \cdot \widetilde{\mathbf{Y}}_{lm}^{S_R}} {\mu_l} \widetilde{\mathbf{Y}}_{lm}^{S_a},
\end{equation}
where 
$\widetilde{\mathbf{Y}}_{lm}^{S_R}$ and $\widetilde{\mathbf{Y}}_{lm}^{S_a}$ are left and right singular vectors, and the corresponding singular values $\mu_l$ are the same as in the continuous case (eq. \ref{eq:mu}).

Since we use the Lebedev quadrature, the dot product $\widetilde{\mathbf{\Phi}} \cdot \widetilde{\mathbf{Y}}_{lm}^{S_R}$ corresponds to exact numerical integration and gives a result identical with the continuous case (eq. \ref{eq:phiY}):
\begin{equation}
    \widetilde{\mathbf{\Phi}} \cdot \widetilde{\mathbf{Y}}_{lm}^{S_R} = \sum_{j=0}^T \Phi_j Y_{lmj}^{S_R} w_j = 
    \sqrt{\frac{4\pi}{2l+1}} \frac{1}{R^{l+1}}Q_{lm}^{mol},
\end{equation}
so the solution to eq. \ref{eq:Ktilde} depends only on the radius $a$ and the multipole moments $Q_{lm}^{mol}$:
\begin{equation} \label{eq:analytical_sigma}
\widetilde{\pmb{\sigma}} =
\sum_{l=0}^{n} \sum_{m=-l}^l \sqrt{ \frac{2l+1}{4\pi} } a^{-l} Q_{lm}^{mol} \widetilde{\mathbf{Y}}_{lm}^{S_a}. 
\end{equation}

The corresponding PC values $q_j$ can be obtained using the quadrature weights $w_j^{S_a}$:
\begin{equation}
  q_i=\sigma_i w^{S_a}_i = \widetilde{\sigma}_i \sqrt{w_i^{S_a}},
\end{equation}
or, in a vector form:
\begin{equation} \label{eq:analytical}
\mathbf{q} = 
\sum_{l=0}^{n} \sum_{m=-l}^l \sqrt{ \frac{2l+1}{4\pi} } a^{-l} Q_{lm}^{mol} \mathbf{Y}_{lm}^{S_a} \odot \mathbf{w}^{S_a},
\end{equation}
where $\mathbf{w}^{S_a}$ is the vector of the quadrature weights for the sphere $S_a$. Therefore, we can use Lebedev grid that shares the origin with a molecule to construct an analytical PC model that exactly reproduces molecular multipole values up to the degree $n$. 

From this model, we can see that the ill-conditioning of the PC fitting due to the decay of the singular values is intrinsic to the inverse electrostatic problem, as the singular values $\mu_l$ decrease with increasing $l$ (eq. \ref{eq:mu}).
Indeed, the higher the multipole moment, the smaller its contribution to the overall electrostatic potential. Also, this contribution  gets smaller as we move the probe further away from the source, and the singular values get smaller with the increasing radius of the probe sphere $R$, or decreasing radius of the source sphere $a$.

The ill-conditioning problems become even more severe as we switch from modeling the MEP using the Lebedev quadrature, which is the best suited to reproduce the molecular multipoles, to an irregular atom-centered quadrature, as shown on a numerical example below.

\begin{figure}[h]
\includegraphics{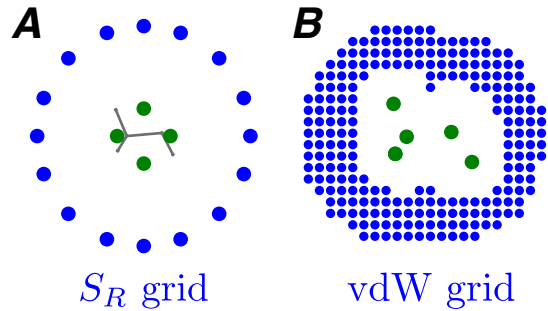}
\caption{\label{fig:models} 
Cross-section representations of the quadratures used for two-sphere model (A) ($n=1$, $N=6$ and $t=11$, $T=194$ for spheres $S_a$ and $S_R$, respectively) as compared with the traditional atom-centered model (B). Green circles correspond to the point charges; blue circles correspond to the reference grid points}
\end{figure}

\section{Lebedev Grid vs. Atom-Centered Model: a Numerical Example}

First, we consider an electrostatic PC model of a methanol molecule with PCs placed at the nodes of the Lebedev quadrature over the sphere $S_a$ ($a=2$ au) (Fig. \ref{fig:models}A). In this case, the PC values can be obtained analytically from the reference multipole moments (eq. \ref{eq:analytical}) or by numerical fitting to the reference MEP over the probe sphere $S_R$ ($R=8$ au, $t=11$, $T=194$):
\begin{equation} \label{eq:numerical}
    \widetilde{\bm{\sigma}} =\sum_{i=1}^{r} \frac{ \mathbf{\widetilde{\Phi}} \cdot \mathbf{\widetilde{v}}_i } {\mu_i} 
    \mathbf{\widetilde{u}}_i,
\end{equation}
where the PC value can be found as $q_j = \widetilde{\sigma}_j \sqrt{w_j^{S_a}}$ and 
the maximum rank $r$ is the number $N$ of quadrature nodes/PCs over the sphere $S_a$.
The quality of the fit is measured using the root mean square deviation (RMSD) calculated over the $T$ nodes of the probe grid:

\begin{equation}
\mathrm{RMSD} = \sqrt{ \frac{\chi^2}{T}  }.
\end{equation}

Naturally, the analytical PC values from eq. \ref{eq:analytical} exactly reproduce the molecular multipole moments up to the degree $n$ defined by the quadrature (Table \ref{tab:multipoles}). 
For each degree $l$ there are $2l+1$ values of order $m$, so overall $(n+1)^2$ multipole moments are reproduced, which matches the dimension $d_n$ of the corresponding basis $\widetilde{\mathbf{Y}}_{S_a}$ (eq. \ref{eq:Ybasis}).
As the dimension $d_n$ increases, more multipole moments are reproduced and the RMSD rapidly approaches zero (Fig. 4 in the supplementary material\cite{SI}).

Since the dimension $d_n$ does not match the number of quadrature nodes $N$ (Table I in the supplementary material\cite{SI}),\cite{Lebedev1976, Sloan1995} we can obtain numerical solutions with eq. \ref{eq:numerical} that are equivalent to the analytical results (eq. \ref{eq:analytical}) by setting the rank $r$ to the dimension of the grid, $d_n=(n+1)^2$ (Table \ref{tab:multipoles}).\cite{f2}

\begin{figure}[t]
\includegraphics{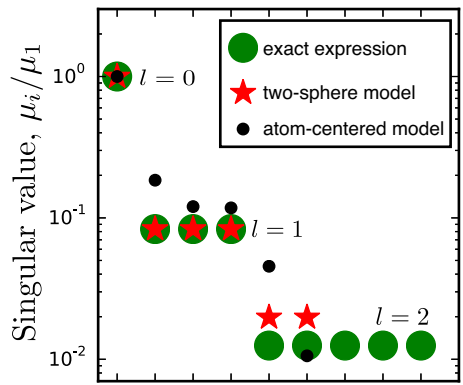}
\caption{\label{fig:singulars}
Normalized singular values $\mu_i/\mu_1$ obtained using the exact analytical expression eq. \ref{eq:mu} (green circles) as compared with the numerical values obtained from SVD of the LS matrix for the two-sphere model (red stars) and for atom-centered model (black circles).
Lebedev quadratures with $n=1$, $N=6$ and $t=11$, $T=194$ were used for the charged $S_a$ ($a=2$ au) and probe $S_R$ ($R=8$ au) spheres, respectively.}
\end{figure}

\begin{figure*}[!htbp]
    \includegraphics{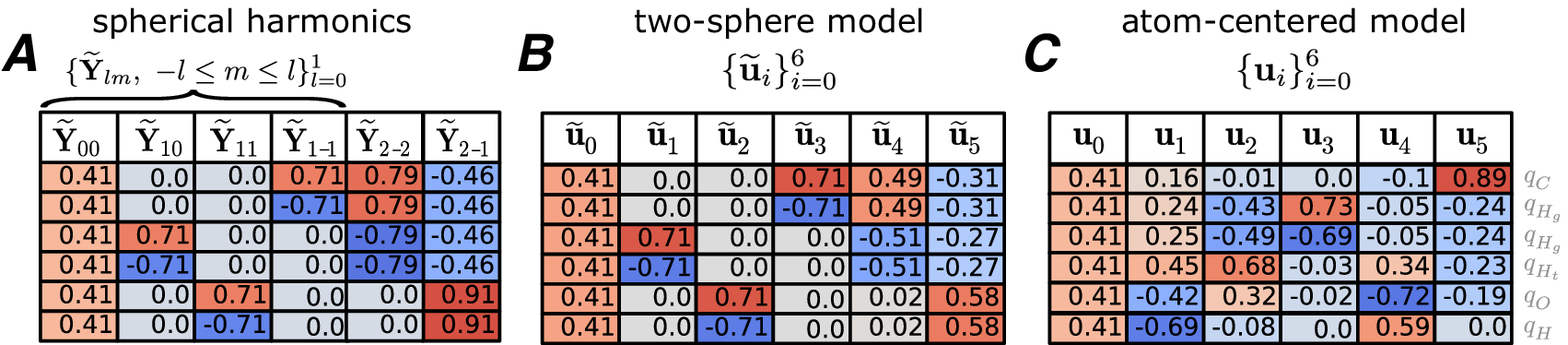}
\caption{\label{fig:bases}
The orthonormal bases of the right singular vectors: basis of spherical harmonics $\widetilde{\mathbf{Y}}_{S_a}$ (A), basis from the numerical SVD of the LS matrix in two-sphere PC model (B), and atom-centered model (C).
}
\end{figure*}

As the first $d_n$ multipole moments $Q_{lm}^{mol}$ are reproduced by the PC model, the first $d_n$ numerical singular values $\mu_i$ exactly match the radius-dependent part (eq. \ref{eq:mu}) from the inverse distance expansion (Fig. \ref{fig:singulars}), and the corresponding right singular vectors $\mathbf{\widetilde{u}}_i$ match the basis $\widetilde{\mathbf{Y}}_{S_a}$ (Fig. \ref{fig:bases}):
\begin{equation}
    \{ \mathbf{\widetilde{u}}_i \}_{i=1}^{d_n} = \{ \widetilde{\mathbf{Y}}_{lm}^{S_a},\ -l\leq m \leq l  \}_{l=0}^{n}.
\end{equation}

If we do not restrict the rank $r$  to the dimension of the grid $d_n$, numerical SVD of the LS matrix $\widetilde{\mathbf{K}}$ (eq. \ref{eq:numerical}) produces $N$ singular vectors/values. While this slightly improves the RMSD (Table \ref{tab:multipoles}), the additional $N-d_n$ singular vectors cannot be described analytically (Fig. \ref{fig:bases}), as they go beyond the dimension $d_n$ of the corresponding basis $\widetilde{\mathbf{Y}}_{S_a}$.
However, in the fortuitous case of the quadrature with $n=1$ and $N=6$, the remaining $6-4=2$ vectors resemble the basis vectors 
$\widetilde{\mathbf{Y}}_{2-2}$ and $\widetilde{\mathbf{Y}}_{2-1}$, so the corresponding quadrupole moments $Q_{2-2}$ and $Q_{2-1}$ are accurately reproduced, although the exact numerical integration of the spherical harmonics $Y_{2-2}$ and $Y_{2-1}$ is not provided by the 6-node Lebedev grid.

Now, we can use the insights from the best-case scenario spherical PC model based on the Lebedev quadrature (Fig. \ref{fig:models}A) to understand the traditional PC fitting problem with atom-centered charges and the probe grid that follows the solvent-accessible surface (vdW grid, Fig. \ref{fig:models}B). From the point of view of the inverse electrostatic model, the atom-centered PC fitting  corresponds to a numerical solution using an irregular and suboptimal integration grid to represent the source charge distribution. This problem can be treated by SVD of the LS matrix $\mathbf{K}$:
\begin{equation} \label{eq:numerical2}
    \mathbf{q} =\sum_{i=1}^{r} \frac{ \mathbf{\Phi} \cdot \mathbf{v}_i } {\mu_i} \mathbf{u}_i,
\end{equation}
where the maximum value of rank $r$ is the number of atoms in the molecule, i.e. $r=6$ in the case of methanol. 

We can see that even in this case the singular vector $\mathbf{u}_1$ with the largest singular value $\mu_1$ corresponds the total charge (Fig. \ref{fig:bases}), which is reproduced with only a slight slight numerical deviation ($<0.01$), a consequence of the molecular charge density spillover beyond the solvent-accessible surface defining the vdW grid.\cite{Ivanov2015,Laio2002}

Although the other singular vectors do not exactly match the corresponding spherical harmonics, the $\mathbf{u}_2$--$\mathbf{u}_4$ vectors can be roughly related to the three components of the dipole moment (Fig. \ref{fig:bases}), and the corresponding singular values are commensurate with the  singular value $\mu_l$ ($l=1$) obtained for the Lebedev grid model (Fig. \ref{fig:singulars}). 
The remaining singular values $\mu_5$ and $\mu_6$ are significantly distorted from the singular value $\mu_l$ ($l=2$), so the components of the quadrupole moment are not reproduced as precisely as the the dipole moment components (Table \ref{tab:multipoles}).

Among all singular vectors $\{ \mathbf{u}_i \}_{i=1}^6$, the singular vector $\mathbf{u}_6$ with the lowest singular value $\mu_6$, which is 100 times smaller than $\mu_1$, is dominated by the contribution from the methyl carbon atom (Fig. \ref{fig:bases}). 
Since such small singular values cause numerical instabilities of the LS solution, once can use a regularization technique such as truncated SVD (tSVD) that reduces the rank $r$ by removing the lowest-$\mu_i$ vector(s) from the SVD expansion.\cite{Hansen2013} 
Removal of $\mathbf{u}_6$ that decreases the rank to $r=5$ leads to dramatic change in the methyl group charges---the carbon atom charge in particular, which drops from $0.22$ to $-0.06$. Yet, these changes lead only to marginal changes in the the multipole moment and RMSD values, a typical example of the buried atom effect (Tables \ref{tab:multipoles}, \ref{tab:atomic_charges}). 
This suggests a natural way to impose a restraint on the buried atom charges without introducing a restraining function into the LS sum $\chi^2$, an addition that can negatively affect the electrostatic properties of the PC model.\cite{Sigfridsson1997, Verstraelen2014}

Further removal of the singular vectors $\mathbf{u}_5$ and $\mathbf{u}_6$ (i.e. $r=4$) leads to severe deterioration of the LS solution,   
as the corresponding multipole moment strongly deviate from the reference values and the RMSD significantly increases (Tables \ref{tab:multipoles}, \ref{tab:atomic_charges}).
Thus, it appears that the tSVD approach should be applied only to the singular vectors that strongly depend on the buried atoms, an important point that will be discussed in detail elsewhere.

\section{The total-charge constraint revisited}

Commonly used PC fitting approaches also modify the LS sum (eq. \ref{eq:chi2}) by adding a Lagrange multiplier $\lambda$ in order to constrain the total charge to the correct value:\cite{Besler1990, Francl1995, Cieplak1995}
\begin{equation}
    \chi^2 \left( \mathbf{q} \right) = |\mathbf{\Phi} - \mathbf{K} \mathbf{q}|^2 + \lambda (\mathbf{1}^\intercal \cdot \mathbf{q} - Q_{0}),
\end{equation}
which increases the dimension of the Hessian matrix 
$\mathbf{H} = \mathbf{K}^\intercal \mathbf{K}$ in the normal equation (eq. \ref{eq:normal}):
\begin{equation} \label{eq:normal2}
    \begin{bmatrix} 
        \mathbf{H} & \mathbf{1} \\
        \mathbf{1}^\intercal & 0 \\
    \end{bmatrix}
    \begin{bmatrix}
        \mathbf{q} \\ \lambda\\
    \end{bmatrix}
    =
    \begin{bmatrix}
        \mathbf{K}^\intercal \mathbf{\Phi} \\ Q_0 \\ 
    \end{bmatrix},
\end{equation}
where $\mathbf{1}$ is an all-ones column-vector.

However, as we have seen, both in the case of the idealized Lebedev grid and the less-than-ideal atom-centered PC models, the Hessian eigenvector with the largest curvature corresponds to the total charge (Fig. \ref{fig:bases} and also Ref. \cite{Ivanov2015}). 
Thus, in the case of the two-sphere PC fitting, the total charge is reproduced exactly ($Q_0<10^{-5}$), while in the atom-centered PC model the total charge only slightly deviates from the exact value due the close proximity of the vdW grid and slight distortion of the total-charge vector $\mathbf{u}_1$ from its analytical analog $\widetilde{\mathbf{Y}}_0$ ($Q_0=0.003$ for methanol, Table \ref{tab:multipoles}).

\begin{table*}[!htbp]
\caption
{\label{tab:multipoles}
Effect of the rank $r$ (eq. \ref{eq:numerical}), degree $n$ (eq. \ref{eq:analytical}) and type of the charge constraint on the methanol multipole moments and the RMSD (kcal/mol) within the Lebedev grid PC model ($a=2$ au, $n=1,2$ and $N=6,14$) with probe sphere $S_R$ ($R=8$ au, $T=194$) and atom-centered PC model with vdW-type grid.
}

\begin{ruledtabular}
\begin{tabular}{ccccccccccccc}
\multirow{1}{*}{PC model, probe grid}
& Details & $Q_{0}$ & $Q_{10}$ & $Q_{11}$ & $Q_{1-1}$ & $Q_{20}$ & $Q_{21}$ & $Q_{2-1}$ & $Q_{22}$ & $Q_{2-2}$ & RMSD \\
\hline
\multirow{3}{*}{$S_a(N=6)$, $S_R (T=194)$}
& eq. \ref{eq:analytical}, $n=1$& 0.000 & 0.000 & -0.325 & -0.565 &  0.000 & 0.000 & 0.000 &  0.000 & 0.000 & 1.762 \\
& SVD, $r=4$ & 0.000 & 0.000 & -0.323 & -0.562 &  0.000 & 0.000 & 0.000 &  0.000 & 0.000 & 1.762 \\
& SVD, $r=6$ & 0.000 & 0.000 & -0.323 & -0.562 & -0.881 & 0.000 & 0.000 & -0.497 & 0.000 & 1.668 \\
\hline
\multirow{3}{*}{$S_a(N=14)$, $S_R (T=194)$}
& eq. \ref{eq:analytical}, $n=2$& 0.000 & 0.000 & -0.325 & -0.565 & -0.887 & 0.000 & 0.000 & -0.503 & 2.882 & 0.559 \\
& SVD, $r=9$ & 0.000 & 0.000 & -0.325 & -0.566 & -0.881 & 0.000 & 0.000 & -0.497 & 2.865 & 0.559 \\
& SVD, $r=14$& 0.000 & 0.000 & -0.325 & -0.566 & -0.881 & 0.000 & 0.000 & -0.497 & 2.865 & 0.425 \\
\hline
\multirow{5}{*}{atom-centered, vdW}
& SVD, $r=6$ & 0.007  & 0.000 & -0.335 & -0.546 & -1.020 & 0.001 & 0.001 & 0.110 & 2.761 & 2.457   \\
& tSVD, $r=5$ & 0.009  & 0.000 & -0.338 & -0.543 & -1.060 & -0.002 & 0.000 & 0.219 & 2.688 & 2.625 \\
& tSVD, $r=4$ & 0.003 & 0.001 & -0.270 & -0.377 & 0.574 & -0.003 & -0.003 & 0.225 & -1.317 & 8.729 \\
& Lagrange, $Q_0 = 0$ 
        & 0.000 & 0.000 & -0.332 & -0.543 & -0.989 & 0.001 & 0.001 & 0.073 & 2.739 & 2.587 \\
& Elimination, $Q_0 = 0$ 
        & 0.000 & 0.000 & -0.332 & -0.543 & -0.989 & 0.001 & 0.001 & 0.073 & 2.739 & 2.587 \\
& SVD, $Q_0 = 0$ 
        & 0.000 & 0.000 & -0.331 & -0.544 & -1.010 & 0.001 & 0.001 & 0.097 & 2.756 & 2.597 \\
& Trivial, $Q_0 = 0$
        & 0.000 & 0.000 & -0.331 & -0.544 & -1.010 & 0.001 & 0.001 & 0.097 & 2.755 & 2.597 \\

\hline
\multicolumn{2}{c}{Reference} 
& 0.000 & 0.000 & -0.325 & -0.565 & -0.887 & 0.000 & 0.000 & -0.503 & 2.882 & --   \\
\end{tabular}
\end{ruledtabular}
\end{table*}

\begin{table*}[!htbp]
\caption
{\label{tab:atomic_charges} 
Effect of the numerical rank $r$ (SVD in eq. \ref{eq:numerical2}) 
and the total-charge constraint
on the values of atom-centered PCs of methanol and the RMSD (kcal/mol).
}
\begin{ruledtabular}
\begin{tabular}{ccccccccccccc}
& & $q_C$ & $q_{H_g}$ & $q_{H_t}$ & $q_O$ & $q_H$ & RMSD & \\
\hline
& SVD, $r=6$ &  0.215 & -0.018 &  0.048 & -0.592 &  0.371 & 2.457 \\
& tSVD, $r=5$ & -0.058 &  0.056 &  0.118 & -0.532 &  0.370 & 2.625 \\
& tSVD, $r=4$ &  0.007 &  0.089 & -0.101 & -0.070 & -0.010 & 8.729  \\
& Lagrange, $Q_0=0$
        &  0.276 & -0.035 &  0.030 & -0.603 &  0.367 & 2.587 \\
& Elimination, $Q_0=0$
        &  0.276 & -0.035 &  0.030 & -0.603 &  0.367 & 2.587 \\
& SVD, $Q_0=0$  
        &  0.214 & -0.019 &  0.047 & -0.593 &  0.370 & 2.597 \\
& Trivial, $Q_0=0$
        & 0.214 &  -0.019 &  0.047 & -0.593 &  0.370 & 2.597 \\
\end{tabular}
\end{ruledtabular}
\end{table*}

Addition of the Lagrange multiplier leads to an extra eigenvector $\mathbf{u}_7$ that appears in the eigenbasis of the Hessian matrix (Table VIII in the supplementary material\cite{SI}).
The curvature along this vector is the smallest in the magnitude ($\kappa_7=-0.009$) and the vector itself primarily depends on the Lagrange multiplier $\lambda$, with only marginal contribution from the PC values.
At the same time, remaining eigenvectors $\{\mathbf{u}\}_{i=1}^6$ preserve the structure of the original eigenbasis, with negligible contribution from the Lagrange multiplier $\lambda$ (Table VIII in the supplementary material\cite{SI}).
Thus, application of the the total charge constraint in addition to already strong restraint (imposed by the eigenvector $\mathbf{u}_1$) appears to be redundant. Moreover, addition of the Lagrange multiplier aggravates the rank deficiency of already ill-conditioned LS problem.\cite{Francl1995, Sigfridsson1997}

Alternatively, the total charge can be constrained by incorporating condition on the proper total charge directly into the LS sum,\cite{Cox1981, Hinsen1997, Jakobsen2014} by eliminating one of the charges and setting it to:
\begin{equation}
	q_n = Q_0^{mol} - \sum_i^{N-1} q_i,
\end{equation}
where $n$ is the index of the eliminated charge. This reduces the dimension of the LS problem by one:
\begin{equation}
	\chi^2 \left( \mathbf{q} \right) = \sum_j^T \left[ \Phi_j - \frac{Q_0^{mol}}{r_{nj}} - \sum_i^{N-1} \left( \frac{1}{r_{ij}} - \frac{1}{r_{nj}} \right) q_i  \right]^2
\end{equation}
and modifies the elements of the Hessian matrix:
\begin{equation}
	H_{km} = \sum_j^{T} \left( \frac{1}{r_{kj}} - \frac{1}{r_{nj}} \right) \left( \frac{1}{r_{mj}} - \frac{1}{r_{nj}} \right).
\end{equation}

Although the solution obtained with this approach is numerically equivalent to the solution with Lagrange multiplier, regardless which atom has been eliminated (Elimination, $Q_0=0$ in Tables \ref{tab:multipoles} and \ref{tab:atomic_charges}), the structure of the right singular vectors becomes disrupted, (Fig. 8 in the supplementary material\cite{SI}) which prevents the application of the truncated SVD to improve the numerical stability of the solution.

Given that even for the atom-centered PC/vdW probe model the total charge value deviates only very slightly from the reference value, it should be possible to correct for this deviation without exacerbating the numerical instabilities of the LS problem, e.g. using the total charge vector $\mathbf{u}_1$. To do that, we convert the SVD solution (eq. \ref{eq:numerical2}) to a system of linear equations:
\begin{equation}\label{eq:svdQ0corr}
    \mathbf{q} = \sum_{i=0}^r \underbrace{\frac{\Phi \cdot \mathbf{v}_i} {\mu_i}}_{c_i} \mathbf{u}_i =
    \mathbf{U} \mathbf{c},
\end{equation}
\begin{equation} \label{ref:sys}
    \mathbf{U}^\intercal \mathbf{q} = \mathbf{c}.
\end{equation}
Then, we replace $\mathbf{u}_1$ in $\mathbf{U}^\intercal$ by an all-ones vector $\mathbf{1}$, and set the corresponding coefficient $c_1$ in $\mathbf{c}$ to the exact value of the molecular total charge $Q_0^{mol}$:
\begin{equation}\label{eq:svdQ0corr1}
    \mathbf{U}_{Q_0}^\intercal \mathbf{q} = \mathbf{c}_{Q_0},
\end{equation}
where 
\begin{equation}
    \mathbf{U}_{Q_0}^\intercal = \begin{bmatrix} \mathbf{1} & \mathbf{u}_2 & \cdots & \mathbf{u}_N \end{bmatrix}^\intercal,
\end{equation}
\begin{equation}\label{eq:svdQ0corr3}
    \mathbf{c}_{Q_0} = \begin{bmatrix} Q_0^{mol} & c_2 & \cdots & c_N \end{bmatrix}^\intercal.
\end{equation}

This approach does not introduce any redundant constraints, preserves the electrostatic properties of the unconstrained solution, and results only in to minor changes in the PC values (SVD, $Q_0=0$ in Tables \ref{tab:multipoles} and \ref{tab:atomic_charges}) and is compatible with truncated SVD.
Also, the error in the total charge value is small enough and can be corrected by simply distributing the $Q_0$ error correction across the atomic charges; this trivial total charge correction gives result nearly identical to eq. \ref{eq:svdQ0corr1} (Trivial, $Q_0=0$ in Tables \ref{tab:multipoles} and \ref{tab:atomic_charges}).

\section{Conclusions}

To understand the origins of the ill-conditioning of the least-squares (LS) point charge (PC) fitting problem, we revisited the PC representation of the molecular electrostatic potential (MEP) from the first principles, as an example of the inverse problem. 

Based on the properties of the Coulomb potential that can be expanded in terms of spherical harmonics, we introduce a model where the MEP of a molecule is exactly reproduced by a charged sphere that has the same multipole moments $Q_{lm}$ as the molecule. Using Lebedev quadrature this continuous model is converted into a discrete PC model, where the PC values are evaluated analytically from the multipole moments $Q_{lm}$ up to the maximum value determined by the quadrature.

In this context, the traditional atom-centered PC model can be viewed as an irregular numerical quadrature, poorly suited to reproduce the multipolar expansion of the MEP. As such, this quadrature only allows integration of the monopole and, approximately, dipole terms. The corresponding large-curvature---or `stiff'\cite{Machta2013,Transtrum2015}---Hessian eigenvectors $\mathbf{u}_i$ can still be related to the corresponding multipoles $Q_{lm}$. This explains previously observed correspondence between the highest-curvature Hessian eigenvectors and the total charge and the dipole moment components;\cite{Ivanov2015,Laio2002} this correspondence quickly breaks down for the higher multipole moments. 

This consideration then reveals the origins of the ill-conditioning of the PC fitting due to the presence of low-curvature---or `sloppy'\cite{Machta2013,Transtrum2015}---vectors  $\mathbf{u}_i$. The intrinsic ill-conditioning arises even in the case of the ideal spherical model: since the higher-rank multipole moments $Q_{lm}$ have smaller contribution to the MEP, the singular values $\mu_l$  decay as $l$ increases. The ill-conditioning is further exacerbated in the numerical treatment of the Lebedev grid model because the number of PCs does not match the dimension of the basis formed by Lebedev quadrature. The remaining singular values/curvatures are even lower in magnitude and do not correspond to particular multipole moments $Q_{lm}$. The same rank-deficiency problems apply to the atom-centered PC grids. However, in that case most of the eigenvectors do not have a direct correspondence to the multipole moments, which leads to even wider spread-out of the singular values/curvatures.

These insights can suggest several ways to alleviate the ill-conditioning of the problem. For instance, the buried atom problem can be addressed by truncating the sloppy singular vectors with dominant contribution from these atom, instead of introducing additional restraining functions\cite{Bayly1993, Dupradeau2010, Zeng2013,Burger2013} that can negatively affect the overall electrostatic properties of the molecule.\cite{Verstraelen2014,Sigfridsson1997} Also, slight deviations of the total charge of the fitted PC solution can be fixed by adjusting the stiff total-charge vector $\mathbf{u}_1$ and the corresponding coordinate $Q_0^{mol}$, rather than introducing a Lagrange multiplier that increases the rank-deficiency of the Hessian matrix.\cite{Francl1995, Sigfridsson1997}

The results presented here can help further application of the PC model in biomolecular simulations. Although the force fields using point charges may not be as accurate as the force fields that explicitly include multipoles and/or polarization effects, the simplicity and computational efficiency of the PC model has ensured its continued survival.\cite{Cerutti2013} In fact, representation of multipoles using the Lebedev grid PC model can provide an alternative to the multipole moment expansion;\cite{Rogers2015} it also can be used to extend recently proposed Distributed Charge Model.\cite{Devereux2014,Gao2014} 

\section{Computational details}
MEP and multipole moments were calculated at the B3LYP/aug-cc-pVDZ level\cite{Lee1988, Becke1993, Stephens1994, Dunning1989} as implemented in Q-Chem package.\cite{QCHEM4}
For atom-centered PC fitting the reference MEP was generated as the cubic grid with linear density 2.8 points/\AA,
followed by the removal of the points outside of 1.0-2.0 van der Waals radii range around each atom (vdW grid).
For the two-sphere PC model the Lebedev quadrature rules were used as implemented in PyQuante package.
\cite{pyquante, pyquante2}
Charge fitting procedures were implemented in the in-house developed \textit{fftoolbox} Python library.\cite{fftoolbox}
SVD was performed using \textit{numpy} library.\cite{numpy} 
Spherical harmonics were accessed from \textit{scipy} library.\cite{scipy}

\begin{acknowledgments}
Q.K.T. is a recipient of the National Science Foundation (NSF) CAREER award CHE-1255641, and Marquette University Way-Klinger Young Scholar Award.
M.V.I. is a recipient of Bournique Memorial Fellowship (Marquette University).
This work used the high-performance computing cluster Père at Marquette University funded by NSF awards OCI-0923037 and CBET-0521602, and the Extreme Science and Engineering Discovery Environment (XSEDE) supported by NSF grant ACI-1053575.
\end{acknowledgments}


\begin{thebibliography}{67}
\bibitem [{f1()}]{f1}%
  \BibitemOpen
  \href@noop {} {}\bibinfo {note} {{For practical purpose we use real-valued
  spherical harmonics, thus to compact the derivation and not to obscure the
  main idea of the work $Y_{lm}$ with $m<0$ ($m>0$) corresponds to $Y_{lms}$
  ($Y_{lmc}$) in standard notation.}}\BibitemShut {Stop}%
\end{thebibliography}
\end{document}